\documentclass{Liebert_Author}

\usepackage{authblk}
\usepackage{amssymb,amsmath,framed}
\usepackage{graphicx}
\usepackage{xcolor}

\usepackage{bm}
\expandafter\ifx\csname package@font\endcsname\relax\else
 \expandafter\expandafter
 \expandafter\usepackage
 \expandafter\expandafter
 \expandafter{\csname package@font\endcsname}
\fi
\def\bq{\begin{equation}}
\def\eq{\end{equation}}
\def\bqy{\begin{eqnarray}}
\def\eqy{\end{eqnarray}}

\title{Dwellers in the Deep: Biological Consequences of Dark Oxygen} 

\author{Manasvi Lingam,$^{1,2,3\ast}$, Amedeo Balbi$^{4}$ and Madhur Tiwari$^{1}$\\
{$^{1}$Department of Aerospace, Physics and Space Sciences, Florida Institute of Technology, Melbourne, FL 32901, USA}\\
{$^{2}$Department of Chemistry and Chemical Engineering, Florida Institute of Technology, Melbourne, FL 32901, USA}\\
{$^{3}$Department of Physics, The University of Texas at Austin, Austin, TX 78712, USA}\\
{$^{4}$Dipartimento di Fisica, Universit\`a di Roma ``Tor Vergata", 00133 Roma, Italy}\\
{$^\ast$To whom correspondence should be addressed;}\\
{E-mail: mlingam@fit.edu}
}
\date{}

\begin{document} 

\maketitle


\begin{abstract}
The striking recent putative detection of ``dark oxygen'' (dark O$_2$) sources on the abyssal ocean floor in the Pacific at $\sim 4$ km depth raises the intriguing scenario that complex (i.e., animal-like) life could exist in underwater environments sans oxygenic photosynthesis. In this work, we thus explore the possible (astro)biological implications of this discovery. From the available data, we roughly estimate the concentration of dissolved O$_2$ and the corresponding O$_2$ partial pressure, as well as the flux of O$_2$ production, associated with dark oxygen sources. Based on these values, we infer that organisms limited by internal diffusion may reach maximal sizes of $\sim 0.1$--$1$ mm in habitats with dark O$_2$, while those with circulatory systems might achieve sizes of $\sim 0.1$--$10$ cm. Optimistically, the estimated dark oxygen flux can potentially support biomass densities up to $\sim 3$--$30$ g m$^{-2}$, perhaps surpassing typical reported densities at similar depths in global deep-sea surveys. Finally, we outline how oceanic settings with dark O$_2$ may facilitate the origin(s) of life via the emergence of electrotrophy. Our findings indicate that complex life fueled by dark oxygen is plausibly capable of inhabiting submarine environments devoid of photosynthesis on Earth, conceivably extending likewise to extraterrestrial locations such as icy worlds with subsurface oceans (e.g., Enceladus and Europa), which are likely common throughout the Universe.
\end{abstract}

\section{Introduction and Motivation}\label{SecIntro}

The non-monotonic increase of molecular oxygen (O$_2$) across Earth's geological history exerted a panoply of profound consequences ranging from the formation of the ozone layer (which partially screens ultraviolet radiation) and new minerals to the creation of novel ecological niches and metabolisms \citep{NL02,DEC14,LRP14,Knoll15,FHJ16,KN17,OPJ17,MA24}. Aerobic metabolic reactions are conventionally documented to yield free energy that is roughly an order of magnitude greater than their anaerobic counterparts for a fixed amount of food \citep{TMC07,KB08}. On the other hand, molecular oxygen can readily abstract electrons from certain biomolecules and enable the formation of reactive oxygen species, thus posing toxicity issues and engendering an ``oxygen paradox'' \citep{Dav95,HG15}.

Drawing partly on the above energy-rich nature of aerobic metabolisms, a variety of classic \citep{JRN59,PC68,BR82,AK85} and modern \citep{CGZ05,KS14,Knoll15,LDP21,MKJ23} studies have advanced and/or elaborated upon the hypothesis that O$_2$ is a prerequisite of sorts for the evolution of complex, motile, macroscopic, multicellular life -- or ``complex'' life for short \citep[cf.][]{AHK11}; we caution, however, that a causal link continues to be debated and analyzed \citep{CME20,BLP21,SBD22}. Despite this ambiguity, it is still tenable that O$_2$ was beneficial for the advent of such complex life \citep{KS14,ML21,MKJ23}, which is implicitly treated in the paper as being near-synonymous with metazoans (i.e., animals), since the latter evince the prior attributes and are potent ecosystem engineers \citep{JLS94,NJB11}.

If we now turn our attention to the provenance of Earth's oxygenated atmosphere and oceans, there are a bevy of factors at play \citep{Knoll15,Catling2017}, but the root cause is the emergence of oxygenic photosynthesis \citep{DEC14,LRP14,FHJ16,LDP21}, which synthesizes molecular oxygen as a metabolic product. In this respect, it should be appreciated that only a small fraction of Earth's oceans are amenable to (oxygenic) photosynthesis. The euphotic zone depth or the compensation depth -- both of which crudely encapsulate the spatial extent wherein photosynthesis (chiefly) takes place -- happen to be $\lesssim 100$ m \citep{SG06,LWK07,Mid19}; evidently much shallower than the average ocean depth of $3.7$ km. Moreover, on temperate rocky planets orbiting M-dwarfs, the most widespread type of stars in the Cosmos, these depths could decline to only a few meters \citep{RLR18,LL20}.

Therefore, significant portions of Earth's habitats -- as well as presumably Earth-like planets in the habitable zones of other stars \citep{KWR93,KRK13} -- do not seem conducive to photosynthesis. The situation is exacerbated if we consider worlds that may have habitable conditions in subsurface oceans under substantial icy shells: notable examples in our solar system are Europa and Enceladus \citep{NP16,HHB19}. This group of worlds are not anticipated to support photosynthesis in their oceans, even if they could otherwise be habitable \citep{GNK99,CH01}. This issue is pertinent because such worlds might be among the most common class of habitable worlds in the Milky Way, namely, up to $\sim 3$ orders of magnitude more prevalent than rocky planets in the habitable zone \citep{SJM21,LBM23}.

On the basis of the preceding paragraphs, it would appear that a multitude of habitable worlds and environments in the Universe do not allow for the production of O$_2$ via oxygenic photosynthesis. The ensuing scarcity of O$_2$ would, \emph{prima facie}, potentially diminish the prospects for complex life. However, this reasoning is rendered invalid if a novel source of O$_2$ not dependent on photosynthesis could be identified. In this context, the remarkable putative discovery of ``\emph{dark oxygen}'' generation at the abyssal seafloor in the ($\sim 4$ km deep) Clarion–Clipperton Zone of the Pacific Ocean \citep{SSJ24} stands out as a conceivable avenue in this regard, perhaps by way of polymetallic nodules.

Thus, by taking inspiration from this ostensibly striking result, we explore possible ramifications of the existence of dark oxygen for the genesis and evolution of complex life in Earth's deep oceans and on other worlds, paralleling the ``follow the oxygen'' strategy espoused in \citet{WSH19}.

\section{Prospects for complex life}
In this section, we delve into the prospects for complex life and generic biomass sustained by dark oxygen at the seafloor.

\subsection{Constraints on dark oxygen availability and biomass}

To assess the consequences of dark oxygen production (DOP) for complex life, we begin with sketching some tentative findings inferred from the empirical data of \citet[pg. 737]{SSJ24}, which will be utilized subsequently.
\begin{itemize}
    \item To loosely estimate the concentration of dissolved O$_2$ (denoted by $\mathcal{C}$) furnished exclusively by means of DOP, we may subtract the final steady-state value of $201$--$819$ $\mu$M from the initial (ambient) O$_2$ abundance of $185.2$ $\mu$M that could likely be supplied by external sources (e.g., current Earth's oxygenated oceans) -- as a consistency check, this ambient concentration is similar to deep-sea measurements of O$_2$ in the Pacific ocean \citep{SMC22} -- which thereupon yields $\mathcal{C} \approx 15.8$--$633.8$ $\mu$M. 
    \item Next, we investigate the following cognate scenario. In lieu of the actual DOP, if we envision an oxygen source/reservoir (e.g., atmosphere) in contact with a water body, as on Earth during its history, we can then ask: what is the O$_2$ partial pressure (labeled by $\mathcal{P}$) that would essentially translate to an effective concentration of $\mathcal{C}$ (stemming from DOP)? To answer this question, we harness Henry's Law for dilute aqueous solutions widely used in physical chemistry \citep[Chapter 4D]{AD17}:
    \begin{equation}\label{Henry}
        \mathcal{C} \approx k_H \mathcal{P},
    \end{equation}
    where $k_H(T)$ is Henry's Law ``constant'' for (dilute) aqueous solutions. In reality, $k_H(T)$ is a function of the ambient temperature $T$, canonically modeled through the Van't Hoff equation for simplicity \citep[Section 2.5]{RS99}. For the case of dissolved O$_2$, we adopt \citep{RS23}:\footnote{\url{https://henrys-law.org/henry/casrn/7782-44-7}}
    \begin{equation}\label{kHtemp}
        k_H(T) \approx 1300\,\mathrm{\mu M}\,\mathrm{bar}^{-1} \exp\left(\frac{1500}{T} - \frac{1500}{298.15\,\mathrm{K}}\right).
    \end{equation}
    On substituting $T = 274.75$ K, akin to the Clarion–Clipperton Zone, in Eq. (\ref{kHtemp}), we obtain $k_H \approx 2 \times 10^3\,\mathrm{\mu M}\,\mathrm{bar}^{-1}$; the external total pressure has a relatively modest effect on the solubility (e.g., \citealt{HPW11,CME20} and \citealt[Chapter 13.4]{AE20}), and is therefore not taken into account. After plugging $k_H$ in Eq. (\ref{Henry}) along with the prior range for $\mathcal{C}$, we end up with $\mathcal{P} \approx 7.9$--$318\,\mathrm{mbar}$, the latter of which is tantamount to $\mathcal{P} \approx 3.7$--$150\%$ PAL (viz., present atmospheric level).
    \item Measurements of abyssal dark oxygen suggest that the local flux (denoted by $\mathcal{F}$) attached to DOP is perhaps $\mathcal{F} \approx 1.7$--$18$ mmol O$_2$ m$^{-2}$ d$^{-1}$.
\end{itemize}
Thus, as per the above inferences and calculations, the lower bound of the potential O$_2$ abundance arising from DOP is estimated to be $\mathcal{C}_\mathrm{min} \approx 15.8$ $\mu$M, or equivalently $\mathcal{P}_\mathrm{min} \approx 7.9\,\mathrm{mbar} \approx 3.7\%$ PAL for the scenario of the oxygen source/reservoir interfacing with water. In some instances, we will employ these conservative choices for dark O$_2$, although we shall mostly avail ourselves of the entire O$_2$ range(s).

Before proceeding onward to carrying out a theoretical analysis of organismal size, we highlight a couple of important features from Earth's biota. A variety of initial analyses indicated that O$_2$ levels of $\sim 1$--$10\%$ PAL may be sufficient for the evolution of early metazoans \citep{BM65,PC76,BR91}, while select publications subsequently revised this threshold downwards to $\gtrsim 0.1\%$ PAL \citep{SHK13}. On comparing this range of proposed values for the desired O$_2$ partial pressure with $\mathcal{P}_\mathrm{min}$, this range apparently overlaps with the latter, hinting that animal-like organisms could possibly exist in environs hosting DOP.

From an empirical standpoint, a plethora of \emph{Polychaete} species are documented in oxygen-scarce benthic regions with O$_2$ concentrations of $< 10$ $\mu$M (e.g., \citealt{LHW91} and \citealt[Table 1]{{SFR13}}), which are distinctly lower than $\mathcal{C}_\mathrm{min}$. Likewise, controlled laboratory experiments conducted on the demosponge \emph{Halichondria panicea} have shown that this animal can survive at oxygen levels of $\lesssim 15$ $\mu$M, or $0.5$--$4\%$ PAL to be accurate \citep{MWJ14}, which are respectively compatible with $\mathcal{C}_\mathrm{min}$ and $\mathcal{P}_\mathrm{min}$. Hence, even on empirical grounds, the O$_2$ abundances attributable to DOP may suffice to sustain metazoan-type organisms. On the other hand, according to modern data, carnivory might be rendered challenging at O$_2$ concentrations of $\lesssim 15$ $\mu$M \citep[pg. 13448]{SFR13}, which is close to $\mathcal{C}_\mathrm{min}$ and should be duly borne in mind.

Another key datum that might be partly adduced in favor of harboring animal-like lifeforms (in abyssal settings with dark oxygen) is that at least one known metazoan, the parasite \emph{Henneguya salminicola} of the family \emph{Myxobolidae} \citep{YAN20}, lacks mitochondrial genomes and the capacity for aerobic respiration. Although a rather extreme outlier -- in the sense of being a parasitic and minuscule animal (sub-mm in size) -- this example nevertheless illustrates that animals do not necessarily require O$_2$ for their survival. 

Finally, let us contemplate the idealized setup where we assume that DOP operates continuously to maintain the flux $\mathcal{F}$, and that all the associated O$_2$ is fully consumed by metazoan-like organisms with no other sinks active. In this setup, we may estimate an upper bound on the biomass area density $\sigma_\mathrm{max}$ on the seafloor by resorting to the following order-of-magnitude approach:
\begin{equation}\label{sigmax}
    \sigma_\mathrm{max} \sim \frac{\mathcal{F}}{\mathcal{B}},
\end{equation}
where $\mathcal{B}$ is the mass-specific aerobic metabolic rate (expressed in terms of the wet tissue mass), chosen to be $\mathcal{B} \sim 0.54$ cm$^3$ O$_2$ g$^{-1}$ h$^{-1}$ from \citet{SHK13} for potential early (or ``primitive'') metazoans. We will now invoke the molar volume -- a quantity of relevance for performing unit conversion \citep[pg. 1213]{HPW11} -- which yields $\mathcal{B} \sim 0.58$ mmol O$_2$ g$^{-1}$ d$^{-1}$.

We caution that the mass-specific metabolic rate exhibits variation across and within taxa, and as a function of body mass of temperature -- the extent is not precisely settled (contrast \citealt{DSG05,AC17,DSG22} with \citealt{MGL08,HMG23}) -- owing to which the results involving $\mathcal{B}$ should be viewed as broadly heuristic estimates. At the same time, however, the salient quantities all scale as $\mathcal{B}^\alpha$; wherein we have $|\alpha| \leq 1$, thereby reflecting a modest dependence on $\mathcal{B}$, and revealing that our conclusions ought not be particularly sensitive to the magnitude of $\mathcal{B}$. This opportune facet should be borne in mind henceforth.

After substituting the preceding values for $\mathcal{B}$ and $\mathcal{F}$ in Eq. (\ref{sigmax}), we arrive at $\sigma_\mathrm{max} \sim 2.9$--$31$ g m$^{-2}$. This optimistic range (i.e., likely embodying the greatest viable biomass density) compares favorably with the observed macrofaunal (wet) biomass density of order $0.01$--$1$ g m$^{-2}$ at an ocean depth of $\sim 4$ km in a global survey \citep[Figure 3]{REM06}, but specialized habitats such as some submarine canyons (e.g., Kaikoura Canyon) comprise an unusually high (wet) biomass density on the order of $1000$ g m$^{-2}$ \citep[pg. 2787]{DSR10}. 

Lastly, an interesting property of abyssal regions of the Clarion–Clipperton Zone is that they contain high megafaunal number densities of order $1$ m$^{-2}$ \citep{AZD16,SBH19,DPB21}, which could be mapped to a biomass density of $\sim 1$--$10$ g m$^{-2}$ for fiducial cm-sized metazoans \citep[Table 1]{SBH19}. This value is loosely similar to $\sigma_\mathrm{max}$, with the proviso that the latter is an optimistic biomass estimate spanning all types of organisms.
 
\subsection{Theoretical diffusion and circulation size limits}

It is well-established that unicellular prokaryote-like organisms may not typically exceed a maximal size imposed by the diffusive uptake of ambient substrates \citep{Den93,TK08}. In fact, this bottleneck can restrict the sizes attainable by the biggest prokaryotes \citep{ALK96,SJ01}, albeit other factors also influence the upper-size thresholds of organisms \citep[e.g.,][]{ABG}, as elaborated hereafter.

We obtain a rough upper bound for the organismal radius ($R_\mathrm{max}$) and size ($L_\mathrm{max} \equiv 2 R_\mathrm{max}$) by balancing the supply rate of O$_2$ through diffusion with the metabolic rate of the spheroidal organism, corresponding to left-hand side (LHS) and right-hand side (RHS) of the following equation:
\begin{equation}
    4 \pi R_\mathrm{max} D \mathcal{C} \sim \frac{4\pi}{3} \rho \mathcal{B} R_\mathrm{max}^3,
\end{equation}
where the LHS constitutes a standard biophysics textbook treatment \citep[Chapter 2]{HCB93}, and the RHS (viz., the whole-organism metabolic rate) is constructed by multiplying the organismal mass with the mass-specific aerobic metabolic rate $\mathcal{B}$. In the above equation, $D$ signifies the diffusion coefficient, and $\rho \approx 1$ g cm$^{-3}$ is the mass density of the organism \citep[pg. 72]{MP16}. On simplifying the above equation, we end up with
\begin{equation}\label{MaxDiffSize}
    L_\mathrm{max} \sim \sqrt{\frac{12 D \mathcal{C}}{\rho \mathcal{B}}},
\end{equation}
for the upper-size bound of a spheroidal organism that acquires O$_2$ via diffusive uptake from the ambient environment.

At normal temperature and pressure (NTP), the diffusion coefficient is $2 \times 10^{-5}$ cm$^2$ s$^{-1}$ for O$_2$ \citep{HB96}. Next, we specify $T \approx 274.75$ K and total pressure of $\sim 400$ bar in the Clarion–Clipperton Zone, enabling us to compute the dynamic viscosity $\eta$ from \citet[Table 1]{SZF05}. Upon combining this data with the Stokes–Einstein equation -- namely, $D \propto T/\eta$ \citep[pg. 212]{MP16} -- we arrive at $D \approx 10^{-5}$ cm$^2$ s$^{-1}$ in this region. On substituting the preceding values in Eq. (\ref{MaxDiffSize}), we obtain $L_\mathrm{max} \sim 0.17$--$1.1$ mm. Hence, we conclude that organisms solely reliant on diffusive uptake of dark O$_2$ might reach sizes up to $0.1$--$1$ mm.


For the upcoming two parts pertaining to maximal size thresholds of idealized representations of multicellular organisms \citep{RMA71}, it is worth recognizing that the original models employed -- which are known to satisfactorily reproduce properties of (modern) Earth's biota -- were formulated as a function of the partial O$_2$ pressure. Thus, given the usage of the effective O$_2$ partial pressure ($\mathcal{P}$) as a proxy, the ensuing results should be carefully interpreted because of possible subtle difference(s) between the dissolved O$_2$ concentration $\mathcal{C}$ at the abyssal seafloor and the scenario delineated previously vis-\`a-vis determining $\mathcal{P}$. Furthermore, as some parameters are not precisely quantified for early metazoans and analogous organisms, we rely on fiducial values in such cases.

In the first group, organisms composed of simple aggregates of cells may be reliant on the diffusion of O$_2$ to all cells in their interior for supporting the organism's metabolic needs \citep{ENH}. For a spheroidal lifeform, this constraint yields a upper-size limit of \citep[pg. 433]{CGZ05}:
\begin{equation}\label{MaxDiffusv2}
     L_\mathrm{max} \sim \sqrt{\frac{6 \mathcal{K}}{\rho \mathcal{B}}\left(\mathcal{P} - \mathcal{P}_\mathrm{crit}\right)},
\end{equation}
where $\mathcal{K}$ is the O$_2$ permeability within the lifeform, and $\mathcal{P}_\mathrm{crit}$ is the minimal (internal) oxygen level required for organismal functioning. An extra factor of $\rho$ appears in the denominator because we work with mass-specific metabolic rate $\mathcal{B}$, and not its volume-specific counterpart. We adopt $\mathcal{K} \sim 2.24 \times 10^{-3}$ cm$^2$ bar$^{-1}$ h$^{-1}$ and $\mathcal{P}_\mathrm{crit} \sim 6.67 \times 10^{-4}$ bar from \citet{SHK13}. On plugging these values along with $\mathcal{P}$, $\mathcal{B}$, and $\rho$ from earlier, we obtain $L_\mathrm{max} \sim 0.13$--$0.89$ mm, implying that the largest organisms constrained by internal diffusion may attain sizes of order $0.1$--$1$ mm in habitats with DOP.

In the second category, the putative spheroidal organisms are presumed to host a circulation system that can distribute O$_2$ throughout their bodies, instead of relying on diffusion. For such lifeforms, the upper bound for the organismal size is expressible as \citep[pg. 424]{CGZ05}:
\begin{equation}\label{MaxCirc}
   L_\mathrm{max} \sim \frac{3 \mathcal{K}}{\rho \mathcal{B} \ell} \left(\mathcal{P} - \mathcal{P}_b\right),
\end{equation}
where $\mathcal{P}_b$ is the average dissolved O$_2$ concentration in the blood, which is modeled as $\mathcal{P}/4$ \citep{RMA71}, and $\ell$ is the mean distance between the organism's surface and circulation system (i.e., the thickness of the epidermal layer). As with Eq. (\ref{MaxDiffusv2}), an extra factor of $\rho$ is included in the denominator for conversion purposes. After setting $\mathcal{B} \sim 0.1$ cm$^3$ O$_2$ g$^{-1}$ h$^{-1}$ for bigger invertebrate-like organisms with lower mass-specific metabolic rates \citep{PMB11} and $\ell \sim 2 \times 10^{-3}$ cm \citep{SHK13} in Eq. (\ref{MaxCirc}), we end up with $L_\mathrm{max} \sim 0.2$--$8$ cm. Hence, DOP might possess the capacity, in principle, to sustain ``megafauna'', to wit, lifeforms exhibiting sizes of $\sim 1$--$10$ cm \citep[pg. 3]{REM06}.

\section{Implications for abiogenesis}
The chief purpose of this paper was to analyze the consequences of DOP for complex life. We will, however, delve briefly into a potential feature identified in \citet{SSJ24}: the authors conjectured that the source of dark oxygen may be seawater electrolysis engendered by polymetallic nodules, which seem to have a potential difference reaching $\Delta \Phi_0 \approx 0.95$ V on their surfaces. We tackle the possible ramifications of this feature for the origin(s) of life, a vast and rapidly expanding field of research \citep{PLL16,DWD19,PAB20,SGS20,ML21}.

In 1961, Peter Mitchell elucidated the mechanism(s) of chemiosmosis \citep{PM61}, which led to the 1978 Nobel Prize in Chemistry \citep{PM79}. This ubiquitous biochemical phenomenon involves the flow of charged particles across a semipermeable membrane to synthesize adenosine triphosphate (ATP) \citep{NF13}. The proton motive force (PMF), denoted by $\Delta p$ (units of V), underpinning chemiosmosis is given by \citep[pg. 44]{NF13}:
\begin{equation}\label{PMF}
    \Delta p = \Delta \phi - \frac{2.3 R T}{F} \Delta(pH),
\end{equation}
where $\Delta \phi$ is the transmembrane electric potential difference, and $\Delta(pH) \equiv pH_\mathrm{in} - pH_\mathrm{out}$ is the pH difference between the interior and exterior of the cell. Note that $R$ represents the gas constant, and $F$ labels the Faraday constant. From inspecting Eq. (\ref{PMF}), it is regulated by multiple variables (ranging from organismal to environmental), owing to which a single value cannot be assigned to the PMF; with that said, a fiducial choice of $\Delta p \sim 0.2 \pm 0.1$ V is compatible with experiments and modeling for sundry species \citep[e.g.,][]{RAR08,SKY11,NF13,MP16}.


Taking inspiration from chemiosmosis and the PMF, the famous alkaline hydrothermal vent (AHV) hypothesis for the origin of life (i.e., abiogenesis) postulates that life emerged in a submarine AHV \citep{MB08,RBB14,SHW16,MJR21,MJR23} by harnessing the naturally occurring redox and proton gradients in these sites \citep{RH97}, which may have translated to a total electrochemical gradient equivalent to $\sim 1$ V as per theoretical analyses and laboratory experiments \citep{RBB14,BAR15,SHW16}. These steep gradients are conjectured to have facilitated carbon fixation, prebiotic synthesis, and abiogenesis \citep{MJR21,SBM23}.

If we compare the reported magnitude of $\Delta \Phi_0$, associated with the nodules, against the above values of the PMF and the putative electrochemical gradient at AHVs, it is evident that the former matches and/or surpasses the latter duo. Therefore, this aspect raises the tantalizing notion that the surfaces of polymetallic nodules might also permit the instantiation of electrochemistry conducive to prebiotic chemistry. However, one crucial divergence is that cells as well as AHVs -- the latter via thin inorganic layers composed of minerals (e.g., iron oxyhydroxides) \citep{MR18} -- evince sharp gradients operational across membranes, whereas these nodules, with typical sizes of $\lesssim 1$--$10$ cm \citep{HKK20}, are not anticipated to have those distinctive properties.

To circumvent the preceding caveat, we underscore that electrical currents have been detected in a variety of submarine hydrothermal vents \citep{YNO13,YNK17}, which can be attributed to a sustained redox potential difference of around $0.5$ V between hydrothermal mineral deposits and the surrounding seawater. This current drives the avenue of electrotrophy, whereby electrons serve as the energy source (i.e., electron donor) for microbial metabolisms \citep{DRL22}. Experiments mimicking the electrical current in hydrothermal vent settings, in conjunction with \emph{in situ} analyses, have shown that microbes (e.g., from genus \emph{Thiomicrorhabdus}) may survive and grow by means of electrotrophy \citep{PAD21,YTK23}. Drawing on these strands, \citet{PSK23} proposed that a prototype of electrotrophy might have sparked abiogenesis, and adumbrated a series of conceivable intermediate steps.

With an electric potential of $\lesssim \Delta \Phi_0$ measured at the surfaces of polymetallic nodules, an electron flow and current is rendered credible, which may in turn pave the way for the emergence of life by way of (proto-)electrotrophy. Moreover, these nodules contain a plethora of metals (e.g., manganese) and minerals (e.g., iron oxyhydroxides) \citep{HKK20}, with strong connections to the origin(s) of life \citep{SKD16,MR18,MVM20}, thence perhaps boosting (directly or otherwise) the significance of polymetallic nodules for abiogenesis.

While the exposition in this section hints at these sites (viz., polymetallic nodules in the abyssal seafloor) comprising intriguing candidates for the nurseries of abiogenesis, further theoretical and empirical investigations of the surface potential difference and the material composition of the nodules, \emph{inter alia}, will be important for gauging their viability; a research program paralleling the modeling and experiments delineated in \citet[Sections 9 \& 12]{MJR21} is arguably warranted.

\section{Discussion and conclusions}

Motivated by the striking ostensible discovery of dark oxygen (O$_2$), we analyzed the possible ramifications of dark O$_2$ for ``complex'' life and the origin(s) of life on Earth and beyond; our major surmises are summarized as follows:
\begin{enumerate}
    \item The maximal sizes attainable by idealized unicellular or multicellular organisms (i.e., constrained by internal or external diffusion processes) for the estimated concentrations of dark O$_2$ may be $\sim 0.1$--$1$ mm.
    \item In contrast, the upper-size bounds of organisms with internal circulation systems for the distribution of oxygen could range between $\sim 0.1$ cm to $\sim 10$ cm, with the latter threshold falling under the umbrella of ``megafauna''.
    \item Under optimistic circumstances, the biomass densities might reach as high as $\sim 3$--$30$ g m$^{-2}$, in principle exceeding the reported macrofaunal densities at depths of $\sim 4$ km in global deep-sea surveys, but perhaps of the same order as the inferred metazoan densities in the abyssal locales of the Clarion–Clipperton Zone.
    \item Polymetallic nodules on the seafloor with a surface potential difference of $\lesssim 1$ V are tentatively considered the provenance of dark oxygen production (DOP). We sketched how this ensuing voltage may conceivably aid the emergence of life by enabling prebiotic synthesis to transpire through the pathway of electrotrophy, where electrons function as the energy source.
\end{enumerate}
In the paper, we have chronicled the limitations and/or unknowns of our overall approach (which has a heuristic slant emphasizing simplicity), where appropriate. 

With regard to the limitations, we have often computed theoretical upper bounds that are not necessarily the characteristic values, and not incorporated factors whose impacts are not clearly resolved (e.g., role of salinity in modulating O$_2$ supply and organismal size). Likewise, it was implicitly posited that DOP and the O$_2$ concentration do not change substantially in time (albeit they can exhibit spatial heterogeneity). This facet patently merits follow-up \emph{in situ} explorations to ascertain the spatiotemporal nature of both dark O$_2$ sources and sinks. As for unknowns, while certain environmental parameters (e.g., temperature and pressure) are reasonably attested, the biological variables should be construed as extrapolations based on available knowledge.

Notwithstanding these caveats, most of our findings are endowed with the key advantage that they are consistent with empirical data (e.g., O$_2$ requirements for metazoans outlined previously). The capacity for possibly harboring macrofauna or megafauna (mentioned above) fueled by dark O$_2$ is roughly compatible with the high biodiversity of such metazoans (e.g., \emph{Xenophyophorea} and \emph{Ophiuroidea}) in the Clarion–Clipperton Zone, including nodule-rich regions \citep{AZD16,GHC17,COH20,RWS23,UMV23}, although this feature does not translate to a causal link. In addition, laboratory experiments entailing the reproduction of polymetallic nodules and DOP may help assess whether organisms and/or biomass densities outside/within the preceding ranges are viable.

Furthermore, as many of the results do not explicitly take the root cause of DOP into account, they could be adapted to study lifeforms and ecosystems supported by alternative channels of DOP. A couple of tenable routes worth highlighting in this realm are as follows:
\begin{itemize}
 \item The production of oxidants on the surface (e.g., powered by radiolysis) and their delivery to the ocean can effectively input O$_2$ to the latter even sans photosynthesis \citep{CH01}. This mechanism is presumably active on icy moons like Europa \citep[Chapter 10]{MA24}, whose $\sim 100$ km deep subsurface ocean is estimated to accrue O$_2$ at supply rates of $\sim 3 \times 10^8 - 3 \times 10^{11}$ mol yr$^{-1}$ \citep{RG10,VHP16}.
 \item Microbial dismutation (e.g., chlorite dismutation) could generate dark O$_2$. This metabolic phenomenon has been recently documented in groundwater samples extracted from multiple aquifers in Canada, where the mean dissolved O$_2$ concentration was determined to be $\sim 16.3$ $\mu$M \citep{RHA23}, which is nearly equal to the minimal O$_2$ abundance ($\mathcal{C}_\mathrm{min}$) employed in our calculations.
\end{itemize}
To round off our preliminary venture into this eclectic subject, we reiterate our prefatory statement that marine habitable settings implausible for photosynthesis, especially on icy worlds with subsurface oceans, are likely widespread in the Universe. Therefore, if DOP is feasible and commonplace on this class of worlds -- whether via seawater electrolysis or the prior two routes -- then our analysis may broadly encapsulate the profound consequences of dark oxygen for the prevalence of abiogenesis, complex multicellularity, and perhaps even technological intelligence \citep{ML21} in the Cosmos.

\section*{Acknowledgments}
TBA.

\bibliographystyle{abbrvnat}
\bibliography{DarkOxygen}

\begin{thebibliography}{113}
\providecommand{\natexlab}[1]{#1}
\providecommand{\url}[1]{\texttt{#1}}
\expandafter\ifx\csname urlstyle\endcsname\relax
  \providecommand{\doi}[1]{doi: #1}\else
  \providecommand{\doi}{doi: \begingroup \urlstyle{rm}\Url}\fi

\bibitem[{Alexander}(1971)]{RMA71}
R.~M. {Alexander}.
\newblock \emph{{Size and Shape}}.
\newblock Number~29 in Studies in Biology. London: Edward Arnold, 1971.

\bibitem[{Amon} et~al.(2016){Amon}, {Ziegler}, {Dahlgren}, {Glover}, {Goineau}, {Gooday}, {Wiklund}, and {Smith}]{AZD16}
D.~J. {Amon}, A.~F. {Ziegler}, T.~G. {Dahlgren}, A.~G. {Glover}, A.~{Goineau}, A.~J. {Gooday}, H.~{Wiklund}, and C.~R. {Smith}.
\newblock {Insights into the abundance and diversity of abyssal megafauna in a polymetallic-nodule region in the eastern Clarion-Clipperton Zone}.
\newblock \emph{Sci. Rep.}, 6:\penalty0 30492, July 2016.
\newblock \doi{10.1038/srep30492}.

\bibitem[{Andersen} et~al.(2016){Andersen}, {Berge}, {Gon{\c{c}}alves}, {Hartvig}, {Heuschele}, {Hylander}, {Jacobsen}, {Lindemann}, {Martens}, {Neuheimer}, {Olsson}, {Palacz}, {Prowe}, {Sainmont}, {Traving}, {Visser}, {Wadhwa}, and {Ki{\o}rboe}]{ABG}
K.~H. {Andersen}, T.~{Berge}, R.~J. {Gon{\c{c}}alves}, M.~{Hartvig}, J.~{Heuschele}, S.~{Hylander}, N.~S. {Jacobsen}, C.~{Lindemann}, E.~A. {Martens}, A.~B. {Neuheimer}, K.~{Olsson}, A.~{Palacz}, A.~E.~F. {Prowe}, J.~{Sainmont}, S.~J. {Traving}, A.~W. {Visser}, N.~{Wadhwa}, and T.~{Ki{\o}rboe}.
\newblock {Characteristic Sizes of Life in the Oceans, from Bacteria to Whales}.
\newblock \emph{Annu. Rev. Mar. Sci.}, 8:\penalty0 217--241, Jan. 2016.
\newblock \doi{10.1146/annurev-marine-122414-034144}.

\bibitem[{Atkins} and {De Paula}(2017)]{AD17}
P.~{Atkins} and J.~{De Paula}.
\newblock \emph{{Elements of Physical Chemistry}}.
\newblock Oxford: Oxford University Press, 2017.

\bibitem[{Averill} and {Eldredge}(2020)]{AE20}
B.~{Averill} and P.~{Eldredge}.
\newblock \emph{{General Chemistry: Principles, Patterns, and Applications}}.
\newblock Boston: FlatWorld, 2.1 edition, 2020.

\bibitem[{Barge} et~al.(2015){Barge}, {Abedian}, {Russell}, {Doloboff}, {Cartwright}, {Kidd}, and {Kanik}]{BAR15}
L.~M. {Barge}, Y.~{Abedian}, M.~J. {Russell}, I.~J. {Doloboff}, J.~H.~E. {Cartwright}, R.~D. {Kidd}, and I.~{Kanik}.
\newblock {From Chemical Gardens to Fuel Cells: Generation of Electrical Potential and Current Across Self-Assembling Iron Mineral Membranes}.
\newblock \emph{Angew. Chem.}, 127\penalty0 (28):\penalty0 8302--8305, July 2015.
\newblock \doi{10.1002/ange.201501663}.

\bibitem[{Berg}(1993)]{HCB93}
H.~C. {Berg}.
\newblock \emph{{Random Walks in Biology}}.
\newblock Princeton: Princeton University Press, 1993.

\bibitem[{Berkner} and {Marshall}(1965)]{BM65}
L.~V. {Berkner} and L.~C. {Marshall}.
\newblock {On the Origin and Rise of Oxygen Concentration in the Earth's Atmosphere.}
\newblock \emph{J. Atmos. Sci.}, 22\penalty0 (3):\penalty0 225--261, May 1965.
\newblock \doi{10.1175/1520-0469(1965)022<0225:OTOARO>2.0.CO;2}.

\bibitem[{Bozdag} et~al.(2021){Bozdag}, {Libby}, {Pineau}, {Reinhard}, and {Ratcliff}]{BLP21}
G.~O. {Bozdag}, E.~{Libby}, R.~{Pineau}, C.~T. {Reinhard}, and W.~C. {Ratcliff}.
\newblock {Oxygen suppression of macroscopic multicellularity}.
\newblock \emph{Nat. Commun.}, 12:\penalty0 2838, Jan. 2021.
\newblock \doi{10.1038/s41467-021-23104-0}.

\bibitem[{Butterfield}(2011)]{NJB11}
N.~J. {Butterfield}.
\newblock {Animals and the invention of the Phanerozoic Earth system}.
\newblock \emph{Trends Ecol. Evol.}, 26\penalty0 (2):\penalty0 81--87, 2011.
\newblock \doi{10.1016/j.tree.2010.11.012}.

\bibitem[{Canfield}(2014)]{DEC14}
D.~E. {Canfield}.
\newblock \emph{{Oxygen: A Four Billion Year History}}.
\newblock Science Essentials. Princeton: Princeton University Press, 2014.

\bibitem[{Catling} and {Kasting}(2017)]{Catling2017}
D.~C. {Catling} and J.~F. {Kasting}.
\newblock \emph{{Atmospheric Evolution on Inhabited and Lifeless Worlds}}.
\newblock Cambridge: Cambridge University Press, 2017.

\bibitem[{Catling} et~al.(2005){Catling}, {Glein}, {Zahnle}, and {McKay}]{CGZ05}
D.~C. {Catling}, C.~R. {Glein}, K.~J. {Zahnle}, and C.~P. {McKay}.
\newblock {Why O$_{2}$ Is Required by Complex Life on Habitable Planets and the Concept of Planetary ``Oxygenation Time''}.
\newblock \emph{Astrobiology}, 5\penalty0 (3):\penalty0 415--438, June 2005.
\newblock \doi{10.1089/ast.2005.5.415}.

\bibitem[{Christodoulou} et~al.(2020){Christodoulou}, {O'Hara}, {Hugall}, {Khodami}, {Rodrigues}, {Hilario}, {Vink}, and {Martinez Arbizu}]{COH20}
M.~{Christodoulou}, T.~{O'Hara}, A.~F. {Hugall}, S.~{Khodami}, C.~F. {Rodrigues}, A.~{Hilario}, A.~{Vink}, and P.~{Martinez Arbizu}.
\newblock {Unexpected high abyssal ophiuroid diversity in polymetallic nodule fields of the northeast Pacific Ocean and implications for conservation}.
\newblock \emph{Biogeosciences}, 17\penalty0 (7):\penalty0 1845--1876, Apr. 2020.
\newblock \doi{10.5194/bg-17-1845-2020}.

\bibitem[{Chyba} and {Hand}(2001)]{CH01}
C.~F. {Chyba} and K.~P. {Hand}.
\newblock {Life Without Photosynthesis}.
\newblock \emph{Science}, 292\penalty0 (5524):\penalty0 2026--2027, 2001.
\newblock \doi{10.1126/science.1060081}.

\bibitem[{Clarke}(2017)]{AC17}
A.~{Clarke}.
\newblock \emph{{Principles of Thermal Ecology: Temperature, Energy and Life}}.
\newblock Oxford: Oxford University Press, 2017.

\bibitem[{Cloud}(1976)]{PC76}
P.~{Cloud}.
\newblock {Beginnings of biospheric evolution and their biogeochemical consequences}.
\newblock \emph{Paleobiology}, 2\penalty0 (4):\penalty0 351--387, Jan. 1976.
\newblock \doi{10.1017/S009483730000498X}.

\bibitem[{Cloud Jr.}(1968)]{PC68}
P.~E. {Cloud Jr.}
\newblock {Atmospheric and Hydrospheric Evolution on the Primitive Earth}.
\newblock \emph{Science}, 160\penalty0 (3829):\penalty0 729--736, May 1968.
\newblock \doi{10.1126/science.160.3829.729}.

\bibitem[{Cole} et~al.(2020){Cole}, {Mills}, {Erwin}, {Sperling}, {Porter}, {Reinhard}, and {Planavsky}]{CME20}
D.~B. {Cole}, D.~B. {Mills}, D.~H. {Erwin}, E.~A. {Sperling}, S.~M. {Porter}, C.~T. {Reinhard}, and N.~J. {Planavsky}.
\newblock {On the co‑evolution of surface oxygen levels and animals}.
\newblock \emph{Geobiology}, 18\penalty0 (3):\penalty0 260--281, May 2020.
\newblock \doi{10.1111/gbi.12382}.

\bibitem[{Davies}(1995)]{Dav95}
K.~J.~A. {Davies}.
\newblock {Oxidative stress: the paradox of aerobic life}.
\newblock \emph{Biochem. Soc. Symp.}, 61:\penalty0 1--31, 1995.
\newblock \doi{10.1042/bss0610001}.

\bibitem[{De Leo} et~al.(2010){De Leo}, {Smith}, {Rowden}, {Bowden}, and {Clark}]{DSR10}
F.~C. {De Leo}, C.~R. {Smith}, A.~A. {Rowden}, D.~A. {Bowden}, and M.~R. {Clark}.
\newblock {Submarine canyons: hotspots of benthic biomass and productivity in the deep sea}.
\newblock \emph{Proc. R. Soc. B}, 277\penalty0 (1695):\penalty0 2783--2792, 2010.
\newblock \doi{10.1098/rspb.2010.0462}.

\bibitem[{Deamer}(2019)]{DWD19}
D.~W. {Deamer}.
\newblock \emph{{Assembling Life: How Can Life Begin on Earth and Other Habitable Planets?}}
\newblock Oxford: Oxford University Press, 2019.

\bibitem[{Denny}(1993)]{Den93}
M.~{Denny}.
\newblock \emph{{Air and Water: The Biology and Physics of Life's Media}}.
\newblock Princeton: Princeton University Press, 1993.

\bibitem[{Durden} et~al.(2021){Durden}, {Putts}, {Bingo}, {Leitner}, {Drazen}, {Gooday}, {Jones}, {Sweetman}, {Washburn}, and {Smith}]{DPB21}
J.~M. {Durden}, M.~{Putts}, S.~{Bingo}, A.~B. {Leitner}, J.~C. {Drazen}, A.~J. {Gooday}, D.~O.~B. {Jones}, A.~K. {Sweetman}, T.~W. {Washburn}, and C.~R. {Smith}.
\newblock {Megafaunal Ecology of the Western Clarion Clipperton Zone}.
\newblock \emph{Front. Mar. Sci.}, 8:\penalty0 671062, 2021.
\newblock \doi{10.3389/fmars.2021.671062}.

\bibitem[{Fischer} et~al.(2016){Fischer}, {Hemp}, and {Johnson}]{FHJ16}
W.~W. {Fischer}, J.~{Hemp}, and J.~E. {Johnson}.
\newblock {Evolution of Oxygenic Photosynthesis}.
\newblock \emph{Annu. Rev. Earth Planet. Sci.}, 44:\penalty0 647--683, June 2016.
\newblock \doi{10.1146/annurev-earth-060313-054810}.

\bibitem[{Gaidos} et~al.(1999){Gaidos}, {Nealson}, and {Kirschvink}]{GNK99}
E.~J. {Gaidos}, K.~H. {Nealson}, and J.~L. {Kirschvink}.
\newblock {Life in Ice-Covered Oceans}.
\newblock \emph{Science}, 284\penalty0 (5420):\penalty0 1631--1633, 1999.
\newblock \doi{10.1126/science.284.5420.1631}.

\bibitem[{Glazier}(2005)]{DSG05}
D.~S. {Glazier}.
\newblock {Beyond the ‘3/4-power law’: variation in the intra- and interspecific scaling of metabolic rate in animals}.
\newblock \emph{Biol. Rev.}, 80\penalty0 (4):\penalty0 611--662, 2005.
\newblock \doi{10.1017/S1464793105006834}.

\bibitem[{Glazier}(2022)]{DSG22}
D.~S. {Glazier}.
\newblock {Variable metabolic scaling breaks the law: from ‘Newtonian’ to ‘Darwinian’ approaches}.
\newblock \emph{Proc. R. Soc. B}, 289\penalty0 (1985):\penalty0 20221605, 2022.
\newblock \doi{10.1098/rspb.2022.1605}.

\bibitem[{Gooday} et~al.(2017){Gooday}, {Holzmann}, {Caulle}, {Goineau}, {Kamenskaya}, {Weber}, and {Pawlowski}]{GHC17}
A.~J. {Gooday}, M.~{Holzmann}, C.~{Caulle}, A.~{Goineau}, O.~{Kamenskaya}, A.~A.~T. {Weber}, and J.~{Pawlowski}.
\newblock {Giant protists (xenophyophores, Foraminifera) are exceptionally diverse in parts of the abyssal eastern Pacific licensed for polymetallic nodule exploration}.
\newblock \emph{Biol. Conserv.}, 207:\penalty0 106--116, Mar. 2017.
\newblock \doi{10.1016/j.biocon.2017.01.006}.

\bibitem[{Greenberg}(2010)]{RG10}
R.~{Greenberg}.
\newblock {Transport Rates of Radiolytic Substances into Europa's Ocean: Implications for the Potential Origin and Maintenance of Life}.
\newblock \emph{Astrobiology}, 10\penalty0 (3):\penalty0 275--283, Apr. 2010.
\newblock \doi{10.1089/ast.2009.0386}.

\bibitem[{Halliwell} and {Gutteridge}(2015)]{HG15}
B.~{Halliwell} and J.~M.~C. {Gutteridge}.
\newblock \emph{{Free Radicals in Biology and Medicine}}.
\newblock Oxford: Oxford University Press, 5th edition, 2015.

\bibitem[{Han} and {Bartels}(1996)]{HB96}
P.~{Han} and D.~M. {Bartels}.
\newblock {Temperature Dependence of Oxygen Diffusion in H$_2$O and D$_2$O}.
\newblock \emph{J. Phys. Chem.}, 100\penalty0 (13):\penalty0 5597--5602, 1996.
\newblock \doi{10.1021/jp952903y}.

\bibitem[{Harvey}(1928)]{ENH}
E.~N. {Harvey}.
\newblock {The oxygen consumption of luminous bacteria}.
\newblock \emph{J. Gen. Physiol.}, 11\penalty0 (5):\penalty0 469--475, 1928.
\newblock \doi{10.1085/jgp.11.5.469}.

\bibitem[{Hein} et~al.(2020){Hein}, {Koschinsky}, and {Kuhn}]{HKK20}
J.~R. {Hein}, A.~{Koschinsky}, and T.~{Kuhn}.
\newblock {Deep-ocean polymetallic nodules as a resource for critical materials}.
\newblock \emph{Nat. Rev. Earth Environ.}, 1\penalty0 (3):\penalty0 158--169, Feb. 2020.
\newblock \doi{10.1038/s43017-020-0027-0}.

\bibitem[{Hendrix} et~al.(2019){Hendrix}, {Hurford}, {Barge}, {Bland}, {Bowman}, {Brinckerhoff}, {Buratti}, {Cable}, {Castillo-Rogez}, {Collins}, {Diniega}, {German}, {Hayes}, {Hoehler}, {Hosseini}, {Howett}, {McEwen}, {Neish}, {Neveu}, {Nordheim}, {Patterson}, {Patthoff}, {Phillips}, {Rhoden}, {Schmidt}, {Singer}, {Soderblom}, and {Vance}]{HHB19}
A.~R. {Hendrix}, T.~A. {Hurford}, L.~M. {Barge}, M.~T. {Bland}, J.~S. {Bowman}, W.~{Brinckerhoff}, B.~J. {Buratti}, M.~L. {Cable}, J.~{Castillo-Rogez}, G.~C. {Collins}, S.~{Diniega}, C.~R. {German}, A.~G. {Hayes}, T.~{Hoehler}, S.~{Hosseini}, C.~J.~A. {Howett}, A.~S. {McEwen}, C.~D. {Neish}, M.~{Neveu}, T.~A. {Nordheim}, G.~W. {Patterson}, D.~A. {Patthoff}, C.~{Phillips}, A.~{Rhoden}, B.~E. {Schmidt}, K.~N. {Singer}, J.~M. {Soderblom}, and S.~D. {Vance}.
\newblock {The NASA Roadmap to Ocean Worlds}.
\newblock \emph{Astrobiology}, 19\penalty0 (1):\penalty0 1--27, Jan. 2019.
\newblock \doi{10.1089/ast.2018.1955}.

\bibitem[{Hoehler} et~al.(2023){Hoehler}, {Mankel}, {Girguis}, {McCollom}, {Kiang}, and {J{\o}rgensen}]{HMG23}
T.~M. {Hoehler}, D.~J. {Mankel}, P.~R. {Girguis}, T.~M. {McCollom}, N.~Y. {Kiang}, and B.~B. {J{\o}rgensen}.
\newblock {The metabolic rate of the biosphere and its components}.
\newblock \emph{Proc. Natl. Acad. Sci.}, 120\penalty0 (25):\penalty0 e2303764120, June 2023.
\newblock \doi{10.1073/pnas.2303764120}.

\bibitem[{Hofmann} et~al.(2011){Hofmann}, {Peltzer}, {Walz}, and {Brewer}]{HPW11}
A.~F. {Hofmann}, E.~T. {Peltzer}, P.~M. {Walz}, and P.~G. {Brewer}.
\newblock {Hypoxia by degrees: Establishing definitions for a changing ocean}.
\newblock \emph{Deep Sea Res. I: Oceanogr. Res.}, 58\penalty0 (12):\penalty0 1212--1226, Dec. 2011.
\newblock \doi{10.1016/j.dsr.2011.09.004}.

\bibitem[{Jones} et~al.(1994){Jones}, {Lawton}, and {Shachak}]{JLS94}
C.~G. {Jones}, J.~H. {Lawton}, and M.~{Shachak}.
\newblock {Organisms as Ecosystem Engineers}.
\newblock \emph{Oikos}, 69\penalty0 (3):\penalty0 373--386, Apr. 1994.
\newblock \doi{10.2307/3545850}.

\bibitem[{Judson}(2017)]{OPJ17}
O.~P. {Judson}.
\newblock {The energy expansions of evolution}.
\newblock \emph{Nat. Ecol. Evol.}, 1\penalty0 (6):\penalty0 0138, 2017.
\newblock \doi{10.1038/s41559-017-0138}.

\bibitem[{Kasting} et~al.(1993){Kasting}, {Whitmire}, and {Reynolds}]{KWR93}
J.~F. {Kasting}, D.~P. {Whitmire}, and R.~T. {Reynolds}.
\newblock {Habitable Zones around Main Sequence Stars}.
\newblock \emph{Icarus}, 101\penalty0 (1):\penalty0 108--128, Jan. 1993.
\newblock \doi{10.1006/icar.1993.1010}.

\bibitem[{Ki{\o}rboe}(2008)]{TK08}
T.~{Ki{\o}rboe}.
\newblock \emph{{A Mechanistic Approach to Plankton Ecology}}.
\newblock Princeton: Princeton University Press, 2008.

\bibitem[{Knoll}(1985)]{AK85}
A.~H. {Knoll}.
\newblock {The precambrian evolution of terrestrial life.}
\newblock In M.~D. {Papagiannis}, editor, \emph{The Search for Extraterrestrial Life: Recent Developments}, volume 112 of \emph{IAU Symposium}, pages 201--211. Dordrecht: Springer, Jan. 1985.
\newblock \doi{10.1007/978-94-009-5462-5_29}.

\bibitem[{Knoll}(2011)]{AHK11}
A.~H. {Knoll}.
\newblock {The Multiple Origins of Complex Multicellularity}.
\newblock \emph{Annu. Rev. Earth Planet. Sci.}, 39:\penalty0 217--239, May 2011.
\newblock \doi{10.1146/annurev.earth.031208.100209}.

\bibitem[{Knoll}(2015)]{Knoll15}
A.~H. {Knoll}.
\newblock \emph{{Life on a Young Planet: The First Three Billion Years of Evolution on Earth}}.
\newblock Princeton Science Library. Princeton University Press, 2nd edition, 2015.

\bibitem[{Knoll} and {Nowak}(2017)]{KN17}
A.~H. {Knoll} and M.~A. {Nowak}.
\newblock {The timetable of evolution}.
\newblock \emph{Sci. Adv.}, 3\penalty0 (5):\penalty0 e1603076, May 2017.
\newblock \doi{10.1126/sciadv.1603076}.

\bibitem[{Knoll} and {Sperling}(2014)]{KS14}
A.~H. {Knoll} and E.~A. {Sperling}.
\newblock {Oxygen and animals in Earth history}.
\newblock \emph{Proc. Natl. Acad. Sci.}, 111\penalty0 (11):\penalty0 3907--3908, Mar. 2014.
\newblock \doi{10.1073/pnas.1401745111}.

\bibitem[{Koch}(1996)]{ALK96}
A.~L. {Koch}.
\newblock {What size should a bacterium be? A question of scale}.
\newblock \emph{Annu. Rev. Microbiol.}, 50\penalty0 (1):\penalty0 317--348, 1996.
\newblock \doi{10.1146/annurev.micro.50.1.317}.

\bibitem[{Koch} and {Britton}(2008)]{KB08}
L.~G. {Koch} and S.~L. {Britton}.
\newblock {Aerobic metabolism underlies complexity and capacity}.
\newblock \emph{J. Physiol.}, 586\penalty0 (1):\penalty0 83--95, 2008.
\newblock \doi{10.1089/ast.2006.0119}.

\bibitem[{Kopparapu} et~al.(2013){Kopparapu}, {Ramirez}, {Kasting}, {Eymet}, {Robinson}, {Mahadevan}, {Terrien}, {Domagal-Goldman}, {Meadows}, and {Deshpande}]{KRK13}
R.~K. {Kopparapu}, R.~{Ramirez}, J.~F. {Kasting}, V.~{Eymet}, T.~D. {Robinson}, S.~{Mahadevan}, R.~C. {Terrien}, S.~{Domagal-Goldman}, V.~{Meadows}, and R.~{Deshpande}.
\newblock {Habitable Zones around Main-sequence Stars: New Estimates}.
\newblock \emph{Astrophys. J.}, 765\penalty0 (2):\penalty0 131, Mar. 2013.
\newblock \doi{10.1088/0004-637X/765/2/131}.

\bibitem[{Lane}(2002)]{NL02}
N.~{Lane}.
\newblock \emph{{Oxygen: The Molecule that Made the World}}.
\newblock Oxford: Oxford University Press, 2002.

\bibitem[{Lee} et~al.(2007){Lee}, {Weidemann}, {Kindle}, {Arnone}, {Carder}, and {Davis}]{LWK07}
Z.~{Lee}, A.~{Weidemann}, J.~{Kindle}, R.~{Arnone}, K.~L. {Carder}, and C.~{Davis}.
\newblock {Euphotic zone depth: Its derivation and implication to ocean-color remote sensing}.
\newblock \emph{J. Geophys. Res. Oceans}, 112\penalty0 (C3):\penalty0 C03009, Mar 2007.
\newblock \doi{10.1029/2006JC003802}.

\bibitem[{Levin} et~al.(1991){Levin}, {Huggett}, and {Wishner}]{LHW91}
L.~A. {Levin}, C.~L. {Huggett}, and K.~F. {Wishner}.
\newblock {Control of deep-sea benthic community structure by oxygen and organic-matter gradients in the eastern Pacific Ocean}.
\newblock \emph{J. Mar. Res.}, 49\penalty0 (4):\penalty0 763--800, 1991.

\bibitem[{Lingam} and {Balbi}(2024)]{MA24}
M.~{Lingam} and A.~{Balbi}.
\newblock \emph{{From Stars to Life: A Quantitative Approach to Astrobiology}}.
\newblock Cambridge: Cambridge University Press, 2024.
\newblock URL \url{https://cambridge.org/9781009411219}.

\bibitem[{Lingam} and {Loeb}(2020)]{LL20}
M.~{Lingam} and A.~{Loeb}.
\newblock {Constraints on Aquatic Photosynthesis for Terrestrial Planets around Other Stars}.
\newblock \emph{Astrophys. J. Lett.}, 889\penalty0 (1):\penalty0 L15, Jan. 2020.
\newblock \doi{10.3847/2041-8213/ab6a14}.

\bibitem[{Lingam} and {Loeb}(2021)]{ML21}
M.~{Lingam} and A.~{Loeb}.
\newblock \emph{{Life in the Cosmos: From Biosignatures to Technosignatures}}.
\newblock Cambridge: Harvard University Press, 2021.

\bibitem[{Lingam} et~al.(2023){Lingam}, {Balbi}, and {Mahajan}]{LBM23}
M.~{Lingam}, A.~{Balbi}, and S.~M. {Mahajan}.
\newblock {A Bayesian Analysis of Technological Intelligence in Land and Oceans}.
\newblock \emph{Astrophys. J.}, 945\penalty0 (1):\penalty0 23, Mar. 2023.
\newblock \doi{10.3847/1538-4357/acb6fa}.

\bibitem[{Lovley}(2022)]{DRL22}
D.~R. {Lovley}.
\newblock {Electrotrophy: Other microbial species, iron, and electrodes as electron donors for microbial respirations}.
\newblock \emph{Bioresour. Technol.}, 345:\penalty0 126553, Feb. 2022.
\newblock \doi{10.1016/j.biortech.2021.126553}.

\bibitem[{Luisi}(2016)]{PLL16}
P.~L. {Luisi}.
\newblock \emph{{The Emergence of Life: From Chemical Origins to Synthetic Biology}}.
\newblock Cambridge: Cambridge University Press, 2nd edition, 2016.

\bibitem[{Lyons} et~al.(2014){Lyons}, {Reinhard}, and {Planavsky}]{LRP14}
T.~W. {Lyons}, C.~T. {Reinhard}, and N.~J. {Planavsky}.
\newblock {The rise of oxygen in Earth's early ocean and atmosphere}.
\newblock \emph{Nature}, 506\penalty0 (7488):\penalty0 307--315, Feb. 2014.
\newblock \doi{10.1038/nature13068}.

\bibitem[{Lyons} et~al.(2021){Lyons}, {Diamond}, {Planavsky}, {Reinhard}, and {Li}]{LDP21}
T.~W. {Lyons}, C.~W. {Diamond}, N.~J. {Planavsky}, C.~T. {Reinhard}, and C.~{Li}.
\newblock {Oxygenation, Life, and the Planetary System during Earth's Middle History: An Overview}.
\newblock \emph{Astrobiology}, 21\penalty0 (8):\penalty0 906--923, Aug. 2021.
\newblock \doi{10.1089/ast.2020.2418}.

\bibitem[{Makarieva} et~al.(2008){Makarieva}, {Gorshkov}, {Li}, {Chown}, {Reich}, and {Gavrilov}]{MGL08}
A.~M. {Makarieva}, V.~G. {Gorshkov}, B.-L. {Li}, S.~L. {Chown}, P.~B. {Reich}, and V.~M. {Gavrilov}.
\newblock {Mean mass-specific metabolic rates are strikingly similar across life's major domains: evidence for life's metabolic optimum}.
\newblock \emph{Proc. Natl. Acad. Sci.}, 105\penalty0 (44):\penalty0 16994--16999, 2008.
\newblock \doi{10.1073/pnas.0802148105}.

\bibitem[{Martin} et~al.(2008){Martin}, {Baross}, {Kelley}, and {Russell}]{MB08}
W.~{Martin}, J.~{Baross}, D.~{Kelley}, and M.~J. {Russell}.
\newblock {Hydrothermal vents and the origin of life}.
\newblock \emph{Nat. Rev. Microbiol.}, 6\penalty0 (11):\penalty0 805--814, 2008.
\newblock \doi{10.1038/nrmicro1991}.

\bibitem[{McCollom}(2007)]{TMC07}
T.~M. {McCollom}.
\newblock {Geochemical Constraints on Sources of Metabolic Energy for Chemolithoautotrophy in Ultramafic-Hosted Deep-Sea Hydrothermal Systems}.
\newblock \emph{Astrobiology}, 7\penalty0 (6):\penalty0 933--950, Dec. 2007.
\newblock \doi{10.1089/ast.2006.0119}.

\bibitem[{Middelburg}(2019)]{Mid19}
J.~J. {Middelburg}.
\newblock \emph{{Marine Carbon Biogeochemistry: A Primer for Earth System Scientists}}.
\newblock Springer Briefs in Earth System Sciences. Cham: Springer, 2019.
\newblock \doi{10.1007/978-3-030-10822-9}.

\bibitem[{Mills} et~al.(2023){Mills}, {Krause}, {Jarvis}, and {Cramer}]{MKJ23}
B.~J.~W. {Mills}, A.~J. {Krause}, I.~{Jarvis}, and B.~D. {Cramer}.
\newblock {Evolution of Atmospheric O$_{2}$ Through the Phanerozoic, Revisited}.
\newblock \emph{Annual Review of Earth and Planetary Sciences}, 51:\penalty0 253--276, May 2023.
\newblock \doi{10.1146/annurev-earth-032320-095425}.

\bibitem[{Mills} et~al.(2014){Mills}, {Ward}, {Jones}, {Sweeten}, {Forth}, {Treusch}, and {Canfield}]{MWJ14}
D.~B. {Mills}, L.~M. {Ward}, C.~{Jones}, B.~{Sweeten}, M.~{Forth}, A.~H. {Treusch}, and D.~E. {Canfield}.
\newblock {Oxygen requirements of the earliest animals}.
\newblock \emph{Proc. Natl. Acad. Sci.}, 111\penalty0 (11):\penalty0 4168--4172, Mar. 2014.
\newblock \doi{10.1073/pnas.1400547111}.

\bibitem[{Milo} and {Phillips}(2016)]{MP16}
R.~{Milo} and R.~{Phillips}.
\newblock \emph{{Cell Biology by the Numbers}}.
\newblock New York: Garland Science, 2016.

\bibitem[{Mitchell}(1961)]{PM61}
P.~{Mitchell}.
\newblock {Coupling of Phosphorylation to Electron and Hydrogen Transfer by a Chemi-Osmotic type of Mechanism}.
\newblock \emph{Nature}, 191\penalty0 (4784):\penalty0 144--148, July 1961.
\newblock \doi{10.1038/191144a0}.

\bibitem[{Mitchell}(1979)]{PM79}
P.~{Mitchell}.
\newblock {Keilin's Respiratory Chain Concept and Its Chemiosmotic Consequences}.
\newblock \emph{Science}, 206\penalty0 (4423):\penalty0 1148--1159, Dec. 1979.
\newblock \doi{10.1126/science.388618}.

\bibitem[{Mojzsis}(2021)]{SJM21}
S.~J. {Mojzsis}.
\newblock {Habitable potentials}.
\newblock \emph{Nat. Astron.}, 5:\penalty0 1083--1085, Nov. 2021.
\newblock \doi{10.1038/s41550-021-01529-3}.

\bibitem[{Muchowska} et~al.(2020){Muchowska}, {Varma}, and {Moran}]{MVM20}
K.~B. {Muchowska}, S.~J. {Varma}, and J.~{Moran}.
\newblock {Nonenzymatic Metabolic Reactions and Life’s Origins}.
\newblock \emph{Chem. Rev.}, 120\penalty0 (15):\penalty0 7708--7744, 2020.
\newblock \doi{10.1021/acs.chemrev.0c00191}.

\bibitem[{Nicholls} and {Ferguson}(2013)]{NF13}
D.~G. {Nicholls} and S.~J. {Ferguson}.
\newblock \emph{{Bioenergetics}}.
\newblock London: Academic Press, 4th edition, 2013.

\bibitem[{Nimmo} and {Pappalardo}(2016)]{NP16}
F.~{Nimmo} and R.~T. {Pappalardo}.
\newblock {Ocean worlds in the outer solar system}.
\newblock \emph{J. Geophys. Res. Planets}, 121\penalty0 (8):\penalty0 1378--1399, Aug. 2016.
\newblock \doi{10.1002/2016JE005081}.

\bibitem[{Nursall}(1959)]{JRN59}
J.~R. {Nursall}.
\newblock {Oxygen as a Prerequisite to the Origin of the Metazoa}.
\newblock \emph{Nature}, 183\penalty0 (4669):\penalty0 1170--1172, Apr. 1959.
\newblock \doi{10.1038/1831170b0}.

\bibitem[{Payne} et~al.(2011){Payne}, {McClain}, {Boyer}, {Brown}, {Finnegan}, {Kowalewski}, {Krause Jr.}, {Lyons}, {McShea}, {Novack-Gottshall}, {Smith}, {Spaeth}, {Stempien}, and {Wang}]{PMB11}
J.~L. {Payne}, C.~R. {McClain}, A.~G. {Boyer}, J.~H. {Brown}, S.~{Finnegan}, M.~{Kowalewski}, R.~A. {Krause Jr.}, S.~K. {Lyons}, D.~W. {McShea}, P.~M. {Novack-Gottshall}, F.~A. {Smith}, P.~{Spaeth}, J.~A. {Stempien}, and S.~C. {Wang}.
\newblock {The evolutionary consequences of oxygenic photosynthesis: a body size perspective}.
\newblock \emph{Photosynth. Res.}, 107\penalty0 (1):\penalty0 37--57, Jan. 2011.
\newblock \doi{10.1007/s11120-010-9593-1}.

\bibitem[{Pillot} et~al.(2021){Pillot}, {Amin Ali}, {Davidson}, {Shintu}, {Godfroy}, {Combet-Blanc}, {Bonin}, and {Liebgott}]{PAD21}
G.~{Pillot}, O.~{Amin Ali}, S.~{Davidson}, L.~{Shintu}, A.~{Godfroy}, Y.~{Combet-Blanc}, P.~{Bonin}, and P.-P. {Liebgott}.
\newblock {Identification of enriched hyperthermophilic microbial communities from a deep-sea hydrothermal vent chimney under electrolithoautotrophic culture conditions}.
\newblock \emph{Sci. Rep.}, 11:\penalty0 14782, July 2021.
\newblock \doi{10.1038/s41598-021-94135-2}.

\bibitem[{Pillot} et~al.(2023){Pillot}, {Santiago}, {Kerzenmacher}, and {Liebgott}]{PSK23}
G.~{Pillot}, {\'O}.~{Santiago}, S.~{Kerzenmacher}, and P.-P. {Liebgott}.
\newblock {Spark of Life: Role of Electrotrophy in the Emergence of Life}.
\newblock \emph{Life}, 13\penalty0 (2):\penalty0 356, Jan. 2023.
\newblock \doi{10.3390/life13020356}.

\bibitem[{Preiner} et~al.(2020){Preiner}, {Asche}, {Becker}, {Betts}, {Boniface}, {Camprubi}, {Chandru}, {Erastova}, {Garg}, {Khawaja}, {Kostyrka}, {Machn{\'e}}, {Moggioli}, {Muchowska}, {Neukirchen}, {Peter}, {Pichlh{\"o}fer}, {Radv{\'a}nyi}, {Rossetto}, {Salditt}, {Schmelling}, {Sousa}, {Tria}, {V{\"o}r{\"o}s}, and {Xavier}]{PAB20}
M.~{Preiner}, S.~{Asche}, S.~{Becker}, H.~C. {Betts}, A.~{Boniface}, E.~{Camprubi}, K.~{Chandru}, V.~{Erastova}, S.~G. {Garg}, N.~{Khawaja}, G.~{Kostyrka}, R.~{Machn{\'e}}, G.~{Moggioli}, K.~B. {Muchowska}, S.~{Neukirchen}, B.~{Peter}, E.~{Pichlh{\"o}fer}, {\'A}.~{Radv{\'a}nyi}, D.~{Rossetto}, A.~{Salditt}, N.~M. {Schmelling}, F.~L. {Sousa}, F.~D.~K. {Tria}, D.~{V{\"o}r{\"o}s}, and J.~C. {Xavier}.
\newblock {The Future of Origin of Life Research: Bridging Decades-Old Divisions}.
\newblock \emph{Life}, 10\penalty0 (3):\penalty0 20, Feb. 2020.
\newblock \doi{10.3390/life10030020}.

\bibitem[{Rabone} et~al.(2023){Rabone}, {Wiethase}, {Simon-Lled{\'o}}, {Emery}, {Jones}, {Dahlgren}, {Bribiesca-Contreras}, {Wiklund}, {Horton}, and {Glover}]{RWS23}
M.~{Rabone}, J.~H. {Wiethase}, E.~{Simon-Lled{\'o}}, A.~M. {Emery}, D.~O.~B. {Jones}, T.~G. {Dahlgren}, G.~{Bribiesca-Contreras}, H.~{Wiklund}, T.~{Horton}, and A.~G. {Glover}.
\newblock {How many metazoan species live in the world’s largest mineral exploration region?}
\newblock \emph{Curr. Biol.}, 33\penalty0 (12):\penalty0 2383--2396, 2023.
\newblock \doi{10.1016/j.cub.2023.04.052}.

\bibitem[{Rao} et~al.(2008){Rao}, {Alonso}, {Rand}, {Dick}, and {Pethe}]{RAR08}
S.~P.~S. {Rao}, S.~{Alonso}, L.~{Rand}, T.~{Dick}, and K.~{Pethe}.
\newblock {The protonmotive force is required for maintaining ATP homeostasis and viability of hypoxic, nonreplicating Mycobacterium tuberculosis}.
\newblock \emph{Proc. Natl. Acad. Sci.}, 105\penalty0 (33):\penalty0 11945--11950, Aug. 2008.
\newblock \doi{10.1073/pnas.0711697105}.

\bibitem[{Rex} et~al.(2006){Rex}, {Etter}, {Morris}, {Crouse}, {McClain}, {Johnson}, {Stuart}, {Deming}, {Thies}, and {Avery}]{REM06}
M.~A. {Rex}, R.~J. {Etter}, J.~S. {Morris}, J.~{Crouse}, C.~R. {McClain}, N.~A. {Johnson}, C.~T. {Stuart}, J.~W. {Deming}, R.~{Thies}, and R.~{Avery}.
\newblock {Global bathymetric patterns of standing stock and body size in the deep-sea benthos}.
\newblock \emph{Mar. Ecol. Prog. Ser.}, 317:\penalty0 1--8, July 2006.
\newblock \doi{10.3354/meps317001}.

\bibitem[{Ritchie} et~al.(2018){Ritchie}, {Larkum}, and {Ribas}]{RLR18}
R.~J. {Ritchie}, A.~W.~D. {Larkum}, and I.~{Ribas}.
\newblock {Could photosynthesis function on Proxima Centauri b?}
\newblock \emph{Int. J. Astrobiol.}, 17\penalty0 (2):\penalty0 147--176, Apr 2018.
\newblock \doi{10.1017/S1473550417000167}.

\bibitem[{Ruff} et~al.(2023){Ruff}, {Humez}, {de Angelis}, {Diao}, {Nightingale}, {Cho}, {Connors}, {Kuloyo}, {Seltzer}, {Bowman}, {Wankel}, {McClain}, {Mayer}, and {Strous}]{RHA23}
S.~E. {Ruff}, P.~{Humez}, I.~H. {de Angelis}, M.~{Diao}, M.~{Nightingale}, S.~{Cho}, L.~{Connors}, O.~O. {Kuloyo}, A.~{Seltzer}, S.~{Bowman}, S.~D. {Wankel}, C.~N. {McClain}, B.~{Mayer}, and M.~{Strous}.
\newblock {Hydrogen and dark oxygen drive microbial productivity in diverse groundwater ecosystems}.
\newblock \emph{Nat. Commun.}, 14:\penalty0 3194, June 2023.
\newblock \doi{10.1038/s41467-023-38523-4}.

\bibitem[{Runnegar}(1982)]{BR82}
B.~{Runnegar}.
\newblock {The Cambrian explosion: Animals or fossils?}
\newblock \emph{J. Geol. Soc. Australia}, 29\penalty0 (3):\penalty0 395--411, Oct. 1982.
\newblock \doi{10.1080/00167618208729222}.

\bibitem[{Runnegar}(1991)]{BR91}
B.~{Runnegar}.
\newblock {Precambrian oxygen levels estimated from the biochemistry and physiology of early eukaryotes}.
\newblock \emph{Palaeogeogr. Palaeoclimatol. Palaeoecol.}, 97\penalty0 (1-2):\penalty0 97--111, Dec. 1991.
\newblock \doi{10.1016/0031-0182(91)90186-U}.

\bibitem[{Russell}(2018)]{MR18}
M.~{Russell}.
\newblock {Green Rust: The Simple Organizing 'Seed' of All Life?}
\newblock \emph{Life}, 8\penalty0 (3):\penalty0 35, Aug. 2018.
\newblock \doi{10.3390/life8030035}.

\bibitem[{Russell}(2021)]{MJR21}
M.~J. {Russell}.
\newblock {The ``Water Problem'' (sic), the Illusory Pond and Life’s Submarine Emergence---A Review}.
\newblock \emph{Life}, 11\penalty0 (5):\penalty0 429, 2021.
\newblock \doi{10.3390/life11050429}.

\bibitem[{Russell}(2023)]{MJR23}
M.~J. {Russell}.
\newblock {A self-sustaining serpentinization mega-engine feeds the fougerite nanoengines implicated in the emergence of guided metabolism}.
\newblock \emph{Front. Microbiol.}, 14:\penalty0 1145915, 2023.
\newblock \doi{10.3389/fmicb.2023.1145915}.

\bibitem[{Russell} and {Hall}(1997)]{RH97}
M.~J. {Russell} and A.~J. {Hall}.
\newblock {The emergence of life from iron monosulphide bubbles at a submarine hydrothermal redox and pH front}.
\newblock \emph{J. Geol. Soc.}, 154\penalty0 (3):\penalty0 377--402, May 1997.
\newblock \doi{10.1144/gsjgs.154.3.0377}.

\bibitem[{Russell} et~al.(2014){Russell}, {Barge}, {Bhartia}, {Bocanegra}, {Bracher}, {Branscomb}, {Kidd}, {McGlynn}, {Meier}, {Nitschke}, {Shibuya}, {Vance}, {White}, and {Kanik}]{RBB14}
M.~J. {Russell}, L.~M. {Barge}, R.~{Bhartia}, D.~{Bocanegra}, P.~J. {Bracher}, E.~{Branscomb}, R.~{Kidd}, S.~{McGlynn}, D.~H. {Meier}, W.~{Nitschke}, T.~{Shibuya}, S.~{Vance}, L.~{White}, and I.~{Kanik}.
\newblock {The Drive to Life on Wet and Icy Worlds}.
\newblock \emph{Astrobiology}, 14\penalty0 (4):\penalty0 308--343, Apr. 2014.
\newblock \doi{10.1089/ast.2013.1110}.

\bibitem[{Sahai} et~al.(2016){Sahai}, {Kaddour}, and {Dalai}]{SKD16}
N.~{Sahai}, H.~{Kaddour}, and P.~{Dalai}.
\newblock {The Transition from Geochemistry to Biogeochemistry}.
\newblock \emph{Elements}, 12\penalty0 (6):\penalty0 389--394, Dec. 2016.
\newblock \doi{10.2113/gselements.12.6.389}.

\bibitem[{Sander}(1999)]{RS99}
R.~{Sander}.
\newblock {Modeling Atmospheric Chemistry: Interactions between Gas-Phase Species and Liquid Cloud/Aerosol Particles}.
\newblock \emph{Surv. Geophys.}, 20\penalty0 (1):\penalty0 1--31, Jan. 1999.
\newblock \doi{10.1023/A:1006501706704}.

\bibitem[{Sander}(2023)]{RS23}
R.~{Sander}.
\newblock {Compilation of Henry's law constants (version 5.0.0) for water as solvent}.
\newblock \emph{Atmos. Chem. Phys.}, 23\penalty0 (19):\penalty0 10901--12440, Oct. 2023.
\newblock \doi{10.5194/acp-23-10901-2023}.

\bibitem[{Sarmiento} and {Gruber}(2006)]{SG06}
J.~L. {Sarmiento} and N.~{Gruber}.
\newblock \emph{{Ocean Biogeochemical Dynamics}}.
\newblock Princeton University Press, 2006.

\bibitem[{Sasselov} et~al.(2020){Sasselov}, {Grotzinger}, and {Sutherland}]{SGS20}
D.~D. {Sasselov}, J.~P. {Grotzinger}, and J.~D. {Sutherland}.
\newblock {The origin of life as a planetary phenomenon}.
\newblock \emph{Sci. Adv.}, 6\penalty0 (6):\penalty0 eaax3419, Feb. 2020.
\newblock \doi{10.1126/sciadv.aax3419}.

\bibitem[{Schmelzer} et~al.(2005){Schmelzer}, {Zanotto}, and {Fokin}]{SZF05}
J.~W.~P. {Schmelzer}, E.~D. {Zanotto}, and V.~M. {Fokin}.
\newblock {Pressure dependence of viscosity}.
\newblock \emph{J. Chem. Phys.}, 122\penalty0 (7):\penalty0 074511--074511, Feb. 2005.
\newblock \doi{10.1063/1.1851510}.

\bibitem[{Schulz} and {J{\o}rgensen}(2001)]{SJ01}
H.~N. {Schulz} and B.~B. {J{\o}rgensen}.
\newblock {Big Bacteria}.
\newblock \emph{Annu. Rev. Microbiol.}, 55\penalty0 (1):\penalty0 105--137, 2001.
\newblock \doi{10.1146/annurev.micro.55.1.105}.

\bibitem[{Schwander} et~al.(2023){Schwander}, {Brabender}, {Mrnjavac}, {Wimmer}, {Preiner}, and {Martin}]{SBM23}
L.~{Schwander}, M.~{Brabender}, N.~{Mrnjavac}, J.~L.~E. {Wimmer}, M.~{Preiner}, and W.~F. {Martin}.
\newblock {Serpentinization as the source of energy, electrons, organics, catalysts, nutrients and pH gradients for the origin of LUCA and life}.
\newblock \emph{Front. Microbiol.}, 14:\penalty0 1257597, 2023.
\newblock \doi{10.3389/fmicb.2023.1257597}.

\bibitem[{Simon-Lled{\'o}} et~al.(2019){Simon-Lled{\'o}}, {Bett}, {Huvenne}, {Schoening}, {Benoist}, {Jeffreys}, {Durden}, and {Jones}]{SBH19}
E.~{Simon-Lled{\'o}}, B.~J. {Bett}, V.~A.~I. {Huvenne}, T.~{Schoening}, N.~M.~A. {Benoist}, R.~M. {Jeffreys}, J.~M. {Durden}, and D.~O.~B. {Jones}.
\newblock {Megafaunal variation in the abyssal landscape of the Clarion Clipperton Zone}.
\newblock \emph{Prog. Oceanogr.}, 170:\penalty0 119--133, Jan. 2019.
\newblock \doi{10.1016/j.pocean.2018.11.003}.

\bibitem[{Smith} et~al.(2022){Smith}, {Messi{\'e}}, {Connolly}, and {Huffard}]{SMC22}
K.~L. {Smith}, M.~{Messi{\'e}}, T.~P. {Connolly}, and C.~L. {Huffard}.
\newblock {Decadal Time-Series Depletion of Dissolved Oxygen at Abyssal Depths in the Northeast Pacific}.
\newblock \emph{Geophys. Res. Lett.}, 49\penalty0 (24):\penalty0 e2022GL101018, Dec. 2022.
\newblock \doi{10.1029/2022GL101018}.

\bibitem[{Soga} et~al.(2011){Soga}, {Kinosita Jr.}, {Yoshida}, and {Suzuki}]{SKY11}
N.~{Soga}, K.~{Kinosita Jr.}, M.~{Yoshida}, and T.~{Suzuki}.
\newblock {Efficient ATP synthesis by thermophilic \emph{Bacillus} F$_o$F$_1$--ATP synthase}.
\newblock \emph{FEBS J.}, 278\penalty0 (15):\penalty0 2647--2654, 2011.
\newblock \doi{10.1111/j.1742-4658.2011.08191.x}.

\bibitem[{Sojo} et~al.(2016){Sojo}, {Herschy}, {Whicher}, {Camprub{\'\i}}, and {Lane}]{SHW16}
V.~{Sojo}, B.~{Herschy}, A.~{Whicher}, E.~{Camprub{\'\i}}, and N.~{Lane}.
\newblock {The Origin of Life in Alkaline Hydrothermal Vents}.
\newblock \emph{Astrobiology}, 16\penalty0 (2):\penalty0 181--197, Feb. 2016.
\newblock \doi{10.1089/ast.2015.1406}.

\bibitem[{Sperling} et~al.(2013{\natexlab{a}}){Sperling}, {Frieder}, {Raman}, {Girguis}, {Levin}, and {Knoll}]{SFR13}
E.~A. {Sperling}, C.~A. {Frieder}, A.~V. {Raman}, P.~R. {Girguis}, L.~A. {Levin}, and A.~H. {Knoll}.
\newblock {Oxygen, ecology, and the Cambrian radiation of animals}.
\newblock \emph{Proc. Natl. Acad. Sci.}, 110\penalty0 (33):\penalty0 13446--13451, Aug. 2013{\natexlab{a}}.
\newblock \doi{10.1073/pnas.1312778110}.

\bibitem[{Sperling} et~al.(2013{\natexlab{b}}){Sperling}, {Halverson}, {Knoll}, {Macdonald}, and {Johnston}]{SHK13}
E.~A. {Sperling}, G.~P. {Halverson}, A.~H. {Knoll}, F.~A. {Macdonald}, and D.~T. {Johnston}.
\newblock {A basin redox transect at the dawn of animal life}.
\newblock \emph{Earth Planet. Sci. Lett.}, 371:\penalty0 143--155, June 2013{\natexlab{b}}.
\newblock \doi{10.1016/j.epsl.2013.04.003}.

\bibitem[{Sperling} et~al.(2022){Sperling}, {Boag}, {Duncan}, {Endriga}, {Marquez}, {Mills}, {Monarrez}, {Sclafani}, {Stockey}, and {Payne}]{SBD22}
E.~A. {Sperling}, T.~H. {Boag}, M.~I. {Duncan}, C.~R. {Endriga}, J.~A. {Marquez}, D.~B. {Mills}, P.~M. {Monarrez}, J.~A. {Sclafani}, R.~G. {Stockey}, and J.~L. {Payne}.
\newblock {Breathless through Time: Oxygen and Animals across Earth’s History}.
\newblock \emph{Biol. Bull.}, 243\penalty0 (2):\penalty0 184--206, 2022.
\newblock \doi{10.1086/721754}.

\bibitem[{Sweetman} et~al.(2024){Sweetman}, {Smith}, {de Jonge}, {Hahn}, {Schroedl}, {Silverstein}, {Andrade}, {Edwards}, {Lough}, {Woulds}, {Homoky}, {Koschinsky}, {Fuchs}, {Kuhn}, {Geiger}, and {Marlow}]{SSJ24}
A.~K. {Sweetman}, A.~J. {Smith}, D.~S.~W. {de Jonge}, T.~{Hahn}, P.~{Schroedl}, M.~{Silverstein}, C.~{Andrade}, R.~L. {Edwards}, A.~J.~M. {Lough}, C.~{Woulds}, W.~B. {Homoky}, A.~{Koschinsky}, S.~{Fuchs}, T.~{Kuhn}, F.~{Geiger}, and J.~J. {Marlow}.
\newblock {Evidence of dark oxygen production at the abyssal seafloor}.
\newblock \emph{Nat. Geosci.}, 17:\penalty0 737--739, 2024.
\newblock \doi{10.1038/s41561-024-01480-8}.

\bibitem[{Uhlenkott} et~al.(2023){Uhlenkott}, {Meyn}, {Vink}, and {Mart{\'\i}nez Arbizu}]{UMV23}
K.~{Uhlenkott}, K.~{Meyn}, A.~{Vink}, and P.~{Mart{\'\i}nez Arbizu}.
\newblock {A review of megafauna diversity and abundance in an exploration area for polymetallic nodules in the eastern part of the Clarion Clipperton Fracture Zone (North East Pacific), and implications for potential future deep-sea mining in this area}.
\newblock \emph{Mar. Biodivers.}, 53\penalty0 (2):\penalty0 22, Apr. 2023.
\newblock \doi{10.1007/s12526-022-01326-9}.

\bibitem[{Vance} et~al.(2016){Vance}, {Hand}, and {Pappalardo}]{VHP16}
S.~D. {Vance}, K.~P. {Hand}, and R.~T. {Pappalardo}.
\newblock {Geophysical controls of chemical disequilibria in Europa}.
\newblock \emph{Geophys. Res. Lett.}, 43\penalty0 (10):\penalty0 4871--4879, May 2016.
\newblock \doi{10.1002/2016GL068547}.

\bibitem[{Ward} et~al.(2019){Ward}, {Stamenkovi{\'c}}, {Hand}, and {Fischer}]{WSH19}
L.~M. {Ward}, V.~{Stamenkovi{\'c}}, K.~{Hand}, and W.~W. {Fischer}.
\newblock {Follow the Oxygen: Comparative Histories of Planetary Oxygenation and Opportunities for Aerobic Life}.
\newblock \emph{Astrobiology}, 19\penalty0 (6):\penalty0 811--824, June 2019.
\newblock \doi{10.1089/ast.2017.1779}.

\bibitem[{Yahalomi} et~al.(2020){Yahalomi}, {Atkinson}, {Neuhof}, {Chang}, {Philippe}, {Cartwright}, {Bartholomew}, and {Huchon}]{YAN20}
D.~{Yahalomi}, S.~D. {Atkinson}, M.~{Neuhof}, E.~S. {Chang}, H.~{Philippe}, P.~{Cartwright}, J.~L. {Bartholomew}, and D.~{Huchon}.
\newblock {A cnidarian parasite of salmon (Myxozoa: Henneguya) lacks a mitochondrial genome}.
\newblock \emph{Proc. Natl. Acad. Sci.}, 117\penalty0 (10):\penalty0 5358--5363, Mar. 2020.
\newblock \doi{10.1073/pnas.1909907117}.

\bibitem[{Yamamoto} et~al.(2013){Yamamoto}, {Nakamura}, {Oguri}, {Kawagucci}, {Suzuki}, {Hashimoto}, and {Takai}]{YNO13}
M.~{Yamamoto}, R.~{Nakamura}, K.~{Oguri}, S.~{Kawagucci}, K.~{Suzuki}, K.~{Hashimoto}, and K.~{Takai}.
\newblock {Generation of Electricity and Illumination by an Environmental Fuel Cell in Deep-Sea Hydrothermal Vents}.
\newblock \emph{Angew. Chem.}, 125\penalty0 (41):\penalty0 10958--10961, Oct. 2013.
\newblock \doi{10.1002/ange.201302704}.

\bibitem[{Yamamoto} et~al.(2017){Yamamoto}, {Nakamura}, {Kasaya}, {Kumagai}, {Suzuki}, and {Takai}]{YNK17}
M.~{Yamamoto}, R.~{Nakamura}, T.~{Kasaya}, H.~{Kumagai}, K.~{Suzuki}, and K.~{Takai}.
\newblock {Spontaneous and Widespread Electricity Generation in Natural Deep-Sea Hydrothermal Fields}.
\newblock \emph{Angew. Chem.}, 129\penalty0 (21):\penalty0 5819--5822, May 2017.
\newblock \doi{10.1002/ange.201701768}.

\bibitem[{Yamamoto} et~al.(2023){Yamamoto}, {Takaki}, {Kashima}, {Tsuda}, {Tanizaki}, {Nakamura}, and {Takai}]{YTK23}
M.~{Yamamoto}, Y.~{Takaki}, H.~{Kashima}, M.~{Tsuda}, A.~{Tanizaki}, R.~{Nakamura}, and K.~{Takai}.
\newblock {In situ electrosynthetic bacterial growth using electricity generated by a deep-sea hydrothermal vent}.
\newblock \emph{ISME J.}, 17\penalty0 (1):\penalty0 12--20, Jan. 2023.
\newblock \doi{10.1038/s41396-022-01316-6}.

\end{thebibliography}

\end{document}